\documentclass[aps,pre,groupedaddress,showpacs]{revtex4-1}
\usepackage{amsmath,amssymb,subfigure,fancybox,txfonts,bm,color,wrapfig}
\usepackage{siunitx,numprint}
\usepackage{graphicx}

\begin{document}

\newcommand{\vc}{\mathbf}
\newcommand{\gvc}[1]{\mbox{\boldmath $#1$}}
\newcommand{\fracd}[2]{\frac{\displaystyle #1}{\displaystyle #2}}
\newcommand{\ave}[1]{\left< #1 \right>}
\newcommand{\red}[1]{\textcolor[named]{Red}{#1}}
\newcommand{\blue}[1]{\textcolor[named]{Blue}{#1}}
\newcommand{\green}[1]{\textcolor[rgb]{0,0.6,0}{#1}}
\newcommand{\del}[3] {\frac{\partial^{#3} #1}{\partial #2^{#3}}}
\newcommand{\dev}[3]{\frac{\text{d}^{#3} #1}{\text{d}#2^{#3}}}
\newcommand{\pdev}[3]{{\text{d}^{#3} #1}/{\text{d}#2^{#3}}}
\newcommand{\pdel}[3]{{\partial^{#3} #1}/{\partial #2^{#3}}}
\newcommand{\intd}[1]{\text{d} {#1}}
\newcommand{\emf}[1]{{\gtfamily \bfseries #1}}
\newcommand{\subti}[1]{\begin{itemize} \item \emf{ #1} \end{itemize}}

\newcommand{\Real}{\operatorname{Re}}
\newcommand{\Imag}{\operatorname{Im}}

\newcommand{\am}{{\bm a}}
\newcommand{\bb}{{\bm b}}
\newcommand{\ff}{{\bm f}}
\newcommand{\pp}{{\bm p}}
\newcommand{\rr}{{\bm r}}
\newcommand{\sm}{{\bm s}}
\newcommand{\tm}{{\bm t}}
\newcommand{\uu}{{\bm u}}
\newcommand{\ww}{{\bm w}}
\newcommand{\xx}{{\bm x}}
\newcommand{\yy}{{\bm y}}
\newcommand{\zz}{{\bm z}}
\newcommand{\Model}{{\mathbb M}}
\newcommand{\RR}{{\mathbb R}}
\newcommand{\NN}{{\mathbb N}}
\newcommand{\oomega}{\mbox{\boldmath $\omega$}}
\newcommand{\WW}{{\bm W}}
\newcommand{\EE}{\mbox{\boldmath $E$}}
\newcommand{\FF}{\mbox{\boldmath $F$}}
\newcommand{\KK}{\mbox{\boldmath $$K$}}
\newcommand{\GG}{\mbox{\boldmath $G$}}
\newcommand{\tr}{\mathrm{T}}
\newcommand{\CC}{\mbox{$\hat{C}$}}
\newcommand{\II}{\mbox{$\hat{I}$}}
\newcommand{\HH}{\mbox{$\hat{H}$}}
\newcommand{\MM}{\mbox{$\hat{M}$}}
\newcommand{\Am}{\mbox{$\hat{A}$}}
\newcommand{\PP}{\mbox{$\hat{P}$}}
\newcommand{\QQ}{\mbox{$\hat{Q}$}}

\title{Optical hyperdimensional soft sensing: Speckle-based touch interface and tactile sensor} 

\author{Kei Kitagawa$^{1}$}
\author{Kohei Tsuji$^{1}$}
\author{Koyo Sagehashi$^{1}$}
\author{Tomoaki Niiyama$^{2}$}
\author{Satoshi Sunada$^{2,4}$}
\email{sunada@se.kanazawa-u.ac.jp}
\affiliation{
$^{1}$Graduate School of Natural Science and Technology, Kanazawa University, Kakuma-machi, Kanazawa, Ishikawa, 920-1192, Japan\\
$^{2}$Faculty of Mechanical Engineering, Institute of Science and Engineering, Kanazawa University, Kakuma-machi, Kanazawa, Ishikawa, 920-1192, Japan\\
}

\begin{abstract}
Hyperdimensional computing (HDC) is an emerging computing paradigm that exploits the distributed representation of input data in a hyperdimensional space, the dimensions of which are typically between 1,000--10,000. 
The hyperdimensional distributed representation enables energy-efficient, low-latency, and noise-robust computations with low-precision and basic arithmetic operations.
In this study, we propose optical hyperdimensional distributed representations based on laser speckles for adaptive, efficient, and low-latency optical sensor processing.
In the proposed approach, sensory information is optically mapped into a hyperdimensional space with $>$250,000 dimensions, enabling HDC-based cognitive processing. 
We use this approach for the processing of a soft-touch interface and a tactile sensor and demonstrate to achieve high accuracy of touch or tactile recognition while significantly reducing training data amount and computational burdens, compared with previous machine-learning-based sensing approaches.  
Furthermore, we show that this approach enables adaptive recalibration to keep high accuracy even under different conditions.
\end{abstract}

\maketitle 

%%%%%%%%%%%%%%%%%%%%%%%%%%  body  %%%%%%%%%%%%%%%%%%%%%%%%%%
\section{Introduction}
Hyperdimensional computing (HDC), also known as vector symbolic architecture, is a brain-inspired computing paradigm \cite{Kanerva:2009aa,10.1145/3538531,10.1145/3558000} that has attracted significant attention driven by global trends in the search for alternatives to the conventional von Neumann computing paradigm. 
HDC takes advantage of the high-dimensional distributed representation of input data. 
For example, an input is represented by a long binary (or bipolar) quasi-random vector, frequently referred to as a {\it hyper vector} (HV), the dimensions of which are typically greater than 1,000 \cite{Kanerva:2009aa}. 
HV representation enables basic arithmetic operations, such as multiplication and addition for HVs, with simple logic operations, such as logical exclusive OR and counters, to build composite HVs that represent objects of interest. 
Because of the hyperdimensional space, two different HVs are likely to be almost orthogonal, which can lead to a holographic representation of input information \cite{Kanerva:2009aa}.  
Another significant advantage of the HV representation is its fault-robust characteristic, that is, the HV representation avoids error-prone bits, such as the most significant bits and the sign bit in conventional binary representation. %
Considering the aforementioned advantages, HDC offers remarkable efficiency as a promising alternative to traditional machine-learning models and has demonstrated its advantages in emerging hardware, such as in-memory computing \cite{Karunaratne:2020aa}. 
HDC-based cognitive processing has been achieved with non-iterative learning without requiring optimization and model turning and applied to various tasks, including language identification \cite{10.1145/2934583.2934624}, speech recognition \cite{8123650}, and object recognition for robotics \cite{Neubert:2019uu}.
It has also been used for analogy-based reasoning \cite{Kanerva_2010_what}, bio-signal processing in electroencephalography \cite{Rahimi:2020vw}, secure computing for the Internet of Things \cite{8814566}, and local processing in adaptive machine learning for wearable devices \cite{Moin:2021aa}.

In-sensor computing (i.e., the local and real-time processing of sensory signals) has the advantage of reducing communication to computational devices and improving latency and security. 
HDC is promising for such in-sensor or near-sensor computing because HDC can encode various types of sensory information as a single HV and enables energy-efficient processing and low latency.
In this study, we considered an HDC-based approach for optical touch-based or tactile sensing. 
In general, such sensors can acquire information regarding force, texture, shape, and temperature through the elastic deformation by a physical contact \cite{TIWANA201217}.
They can be extended to various applications, such as robot interaction, medical probes, and haptic interfaces.

As opposed to electronics-based sensors, optical-sensing approaches exhibit remarkable features, such as high sensitivity, remote access, and immunity to electromagnetic interference. 
There are many optical-sensing approaches, including stretchable waveguide- \cite{Zhaoeaai7529,Kimeabc6878} or fiber-based sensors \cite{Xu:18_OpticsLetter}, 
and speckle-based sensors \cite{FUJIWARA2023100345,FUJIWARA2017677,Shimadera:2022aa}.  
Among them, the speckle-based sensing approach enables highly sensitive detection, combined with an image correlation technique \cite {FUJIWARA2023100345,FUJIWARA2017677} or deep learning \cite{Shimadera:2022aa, Smith:22}. 
However, the image correlation technique suffers from the limitation of vulnerability to noise and lack of dynamic range \cite{ Smith:22}. 
Deep learning-based techniques can achieve higher accuracy and consistency with a wider dynamic range \cite{Smith:22} and multimodal sensing \cite{Shimadera:2022aa}. 
However, they normally require substantial training samples and suffer from the computational burden and adaptivity to environmental changes.

In this study, we employ the HDC concept for the efficient, adaptive, accurate processing of optical sensing data with low computational burden. 
The implementation of the HDC is based on an optical encoding for HV generation. 
Various types of sensory information can be encoded as a speckle pattern, which is used as a significantly long HV (typically with more than 250,000 dimensions).
The high-dimensional distributed representation can be achieved naturally through optical scattering without additional computational burden.
Similar speckle-based high-dimensional mapping techniques have already been utilized in extreme-learning machines and reservoir computing \cite{Sunada:20,Sunada:21,Paudel:20,PhysRevX.10.041037}; however, their application to HDC has not been reported to date. 
The proposed HV generation approach enables accurate cognitive processing based on a straightforward and low-precision operation without an iterative training process using a large amount of training data, which is different from traditional deep learning approaches. 
We apply the proposed approach to a soft touch sensor and a tactile sensor and demonstrate the features of the proposed approach.
This approach paves the way for an optical in-sensor computing paradigm that seamlessly integrates optical-sensing capabilities and information processing.

\section{Principle and methods}
\subsection{Classification using HDC}
\begin{figure}[ht!]
\centering\includegraphics[width=10cm]{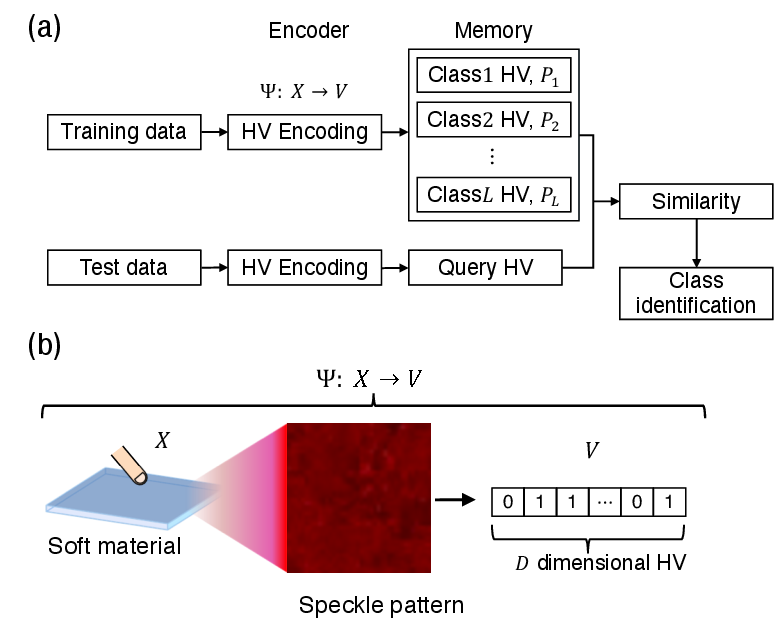}
\caption{\label{fig1} (a) Classification using HDC. 
(b) Optical HV generation scheme using the speckle-based distributed representation of input sensory information.
The touch sensory information is mapped into a speckle pattern, which is used to generate HVs. 
}
\end{figure}

Here, we briefly describe the use of HDC in cognitive processing tasks, such as classification \cite{9107175}. 
HDC consists of an encoder and a memory and can achieve classification in a straightforward manner [Fig.~\ref{fig1}(a)]. 
In HDC, the encoder is used to transform input data into HVs, while the memory is used to store and process the HVs. 
In general, HVs are represented as binary or bipolar vectors with $D$ dimensions, which are chosen independently from $\{0,1\}^D$ or $\{-1,1\}^D$. 
The probability of bit 1 is 0.5. 
Thus, the generated HVs are nearly orthogonal to each other.% for $D \gg 1$. 

Let $\{X_i,l_i\}_{i=1}^N$ be a training dataset with $N$ samples, where $X_i$ and $l_i$ are the $i$-th input vector and target label, respectively.
The number of the target labels is defined as $L$.
$X_i$ is encoded as HV $V_i$ through the encoding function $\Psi: X_i \rightarrow V_i$.
Then, let $P_l$ be an HV representing class $l$, called a {\it prototype vector}.  
Let $\{V_k^l\}_{k=1}^{N_s}$ be a set of the HVs belonging to class $l$, where $k$ and $N_s$ are the sample index and the number of the HVs, respectively.
For the sake of simplicity, we assumed $N_s = N/L$.
The prototype vector $P_l$ for class $l$ is generated with point-wise addition as follows:
\begin{eqnarray}
P_l = \left[V_1^l + V_2^l + \cdots + V_{N_s}^l\right], \label{eq_1}
\end{eqnarray}
where $\left[\cdot\right]$ is the binarization operation used to transform any $D$-dimensional vector into a $D$-dimensional binary vector based on the majority rule.
In this study, we used a simple majority rule, in which ``0''(``1'') is taken if the number of ``0''(``1'') is larger. 
The bias of adding an even number of HVs can be reduced by adding an extra random vector \cite{10.1145/3314326}; however, this bias problem was not addressed in this study. 
All trained prototype vectors, $\{P_1, P_2, \cdots, P_L\}$, are stored in a memory. 

During the inference phase, unknown data can be classified as follows. 
First, the data value is mapped to a high-dimensional space using the same encoding scheme.
This HV is called the {\it query vector} $V_q$. 
Then, the similarity between the generated query vector and all stored prototype vectors is measured.
For any two binary HVs, $A = (A_1, \cdots, A_D)$ and $B = (B_1, \cdots, B_D)$, the similarity can be measured using the Hamming distance, which is given by 
\begin{eqnarray}
\mbox{Ham}(A,B) = \sum_{j=1}^D A_j \oplus B_j \label{eq_2}, %1_{A_j \ne B_j}. 
\end{eqnarray}
where $\oplus$ is the XOR operation, which is unity if and only if arguments $A_j$ and $B_j$ differ; otherwise, it is zero. 
$\mbox{Ham}(A,B) = 0$ only for $A = B$, whereas $\mbox{Ham}(A,B) \approx 0.5D$ if $A$ and $B$ are nearly orthogonal or dissimilar. 
Finally, the unknown data are classified into class $l^*$, with which it has the highest similarity, that is, the shortest Hamming distance, as follows:
\begin{eqnarray}
l^* = \mbox{argmin}_{l} \mbox{Ham}(P_l,V_q). \label{eq_3}
\end{eqnarray}

\subsection{Optical hyperdimensional mapping for generating HVs}
The core of HDC is its encoding process to map the input data into HVs. 
In conventional HDC, encoding methods are based on mathematical operations \cite{9107175}. 
For example, a record-based encoding scheme utilizes two types of HVs, representing the feature position and feature value \cite{9107175}. 
For the position encoding of a feature vector with $m$ elements, $m$ HVs are randomly generated. 
Then, each feature value is discretized to $n$ levels.
The level values are represented by $n$ HVs, which in turn are generated such that HVs of neighbor levels are correlated.
Details of this operation and various other encoding schemes can be found in the literature \cite{aygun2023learning}. 
These standard encoding schemes require arithmetic operations, which are typically computationally expensive. 
In contrast to previous work, we did not use arithmetic operations in this study; rather, an optical scattering process was used for HV generation. 

The proposed HV generation is based on optical encoding using the modulation sensitivity of speckle patterns in soft materials [Fig.~\ref{fig1}(b)]. 
As is well known, optical scattering in a diffusive material is highly sensitive to external stimuli to the material \cite{Goodman:76,goodman2007speckle}. 
Therefore, material deformation created by touch interactions is encoded as a speckle pattern.
Thus, as opposed to conventional approaches, some arithmetic operations are skipped, and the memory for HV storage is not required in our optical approach. 

\begin{figure}[ht!]
\centering\includegraphics[width=13cm]{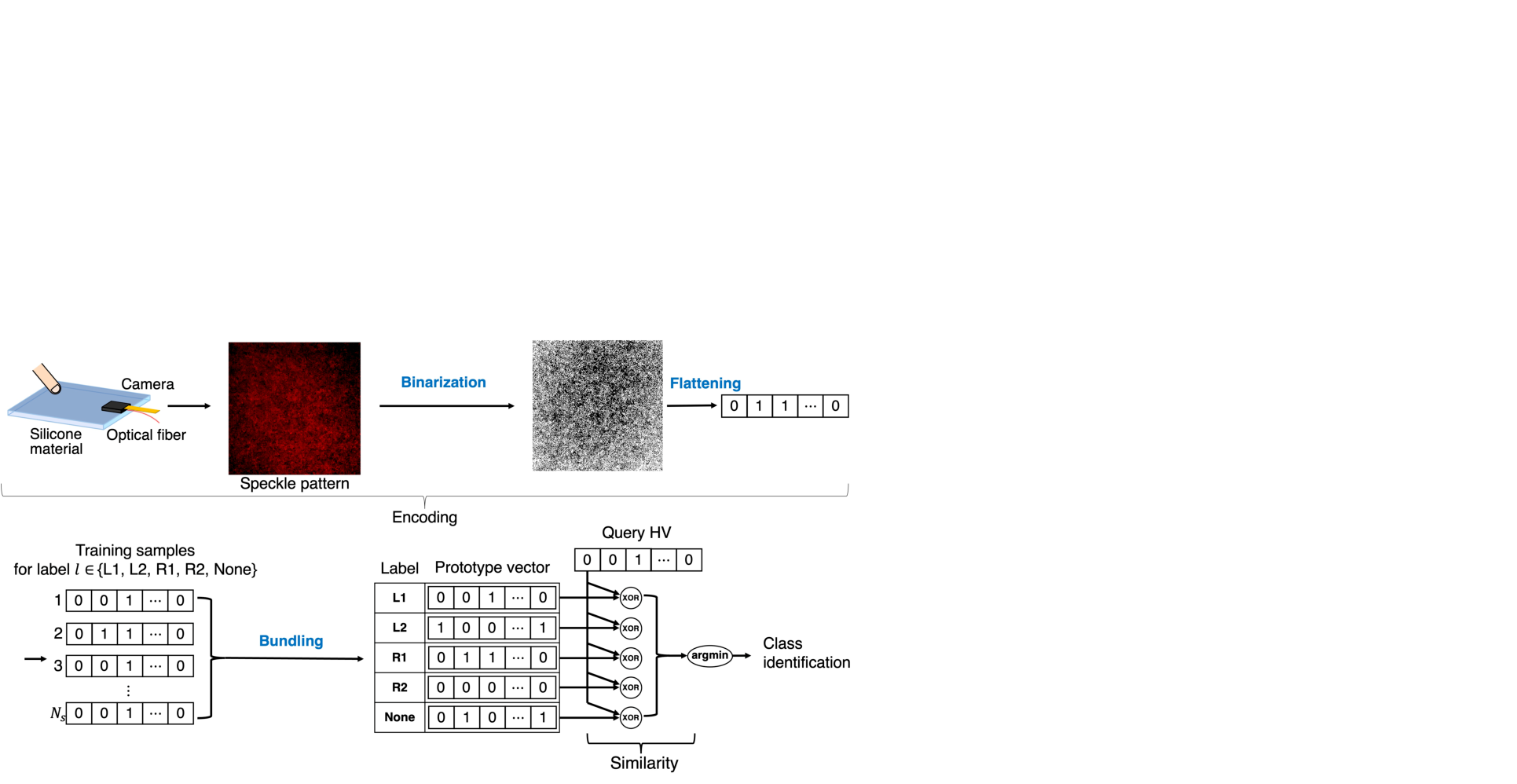}
\caption{\label{fig2} Overview of the HDC-based classification approach 
for optical soft sensors.}
\end{figure}

The sensing of touch stimuli is treated as a classification problem using HDC (Fig.~\ref{fig2}). 
To classify the stimulus information, speckle images are captured using a camera and then used as feature HVs.
In our approach, the feature HVs are thresholded and binarized such that ``0'' and ``1'' appear with the same probability of 0.5. 
Within the same label, HVs are added to generate a prototype vector [see Eq.~(\ref{eq_1})].
The prototype vector contains the information features of each class. 
The test data, which are also obtained as speckle patterns, are mapped to a query HV and classified using the similarity measurement of each prototype vector [Eq.~(\ref{eq_3})]. 

A remarkable feature of the optical approach is the natural generation of HVs with more than 100,000 dimensions using a simple optical setup. 
The dimensions of the generated HVs depend on the number of pixels in the image sensor used to measure the speckle patterns. 
High dimensionality is important for the orthogonality between different HVs and a reliable symbolic representation. 
However, longer HVs require more memory for the storage; therefore, 
they should be shorter. 
Typically, existing HDC approaches use HVs with 1,000--10,000 dimensions and require them to be stored for each input value. 
Our approach can generate HVs directly from an external stimulus; thus, a significant memory usage reduction is expected.

\section{Results}
\subsection{Optical touch sensor}
\subsubsection{Setup}
\begin{figure}[ht!]
\centering\includegraphics[width=13cm]{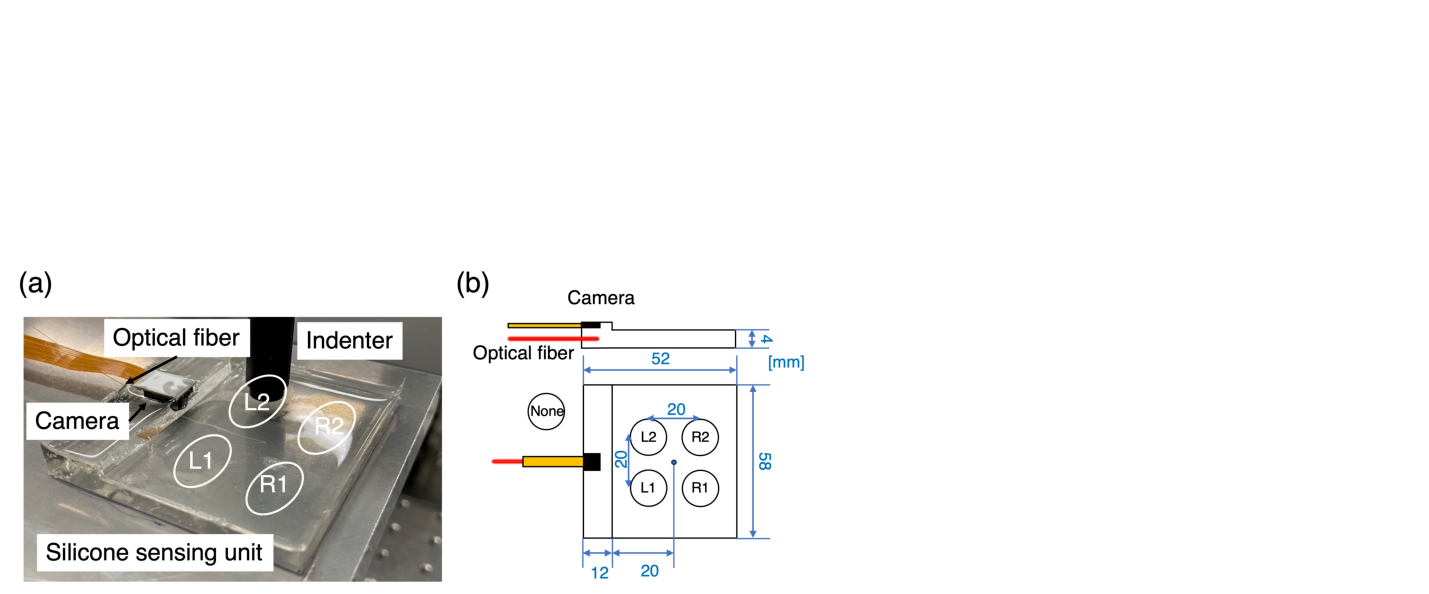}
\caption{\label{fig3} 
(a) Optical touch sensor. 
(b) Schematic of the touch sensor. 
Contact positions can be identified using the HDC-based approach. 
}
\end{figure}

Figure~\ref{fig3}(a) shows an optical touch sensor used for an interface device. 
The sensing part consists of a transparent silicone elastomer, which is coupled with an optical fiber and a compact camera. 
Laser light is scattered inside the silicone and forms speckle patterns owing to the complex scattering process. 
The speckle patterns are captured using the camera. 
(See Supplement 1, Sections 1A--1C, for the details.)
Considering that the speckle patterns change depending on the contact with the surface of the sensing part, the information concerning the contact action can be identified by learning the change characteristics. 
We used the HDC-based approach to identify the contact positions in the sensing part. 
As a proof-of-concept, we demonstrated the identification of contact positions, labeled as L1, L2, R1, R2, and None [Fig.~\ref{fig3}(b)]. 
``None'' represents no contact with the sensing part. 
In this experiment, the training data were collected automatically using a robotic arm.

The silicone surface was pushed at the positions labeled L1, L2, R1, and R2 
by a solid indenter mounted on the robotic arm.
The contact positions (L1, L2, R1, and R2) were randomly chosen.
The pushing depth was estimated as $<$ 2 mm.
The resulting speckle patterns were measured using the camera. 
The temperature was approximately 23.3 $\pm$ 0.2 $^\circ$C. 
Data samples $\{X_i,l_i\}_{i=1}^{N_T}$ were collected, where $X_i$ and 
$l_i \in \{\mbox{L1, L2, R1, R2, None}\}$ represent the $i$-th speckle pattern and the label of contact position, respectively. 
$N_T = 500$ is the total number of collected data samples, and 80 $\%$ of the data samples were used for training (i.e., prototype vector generation). 

\subsubsection{Speckle, HVs, and prototype vectors}
\begin{figure}[ht!]
\centering\includegraphics[width=13cm]{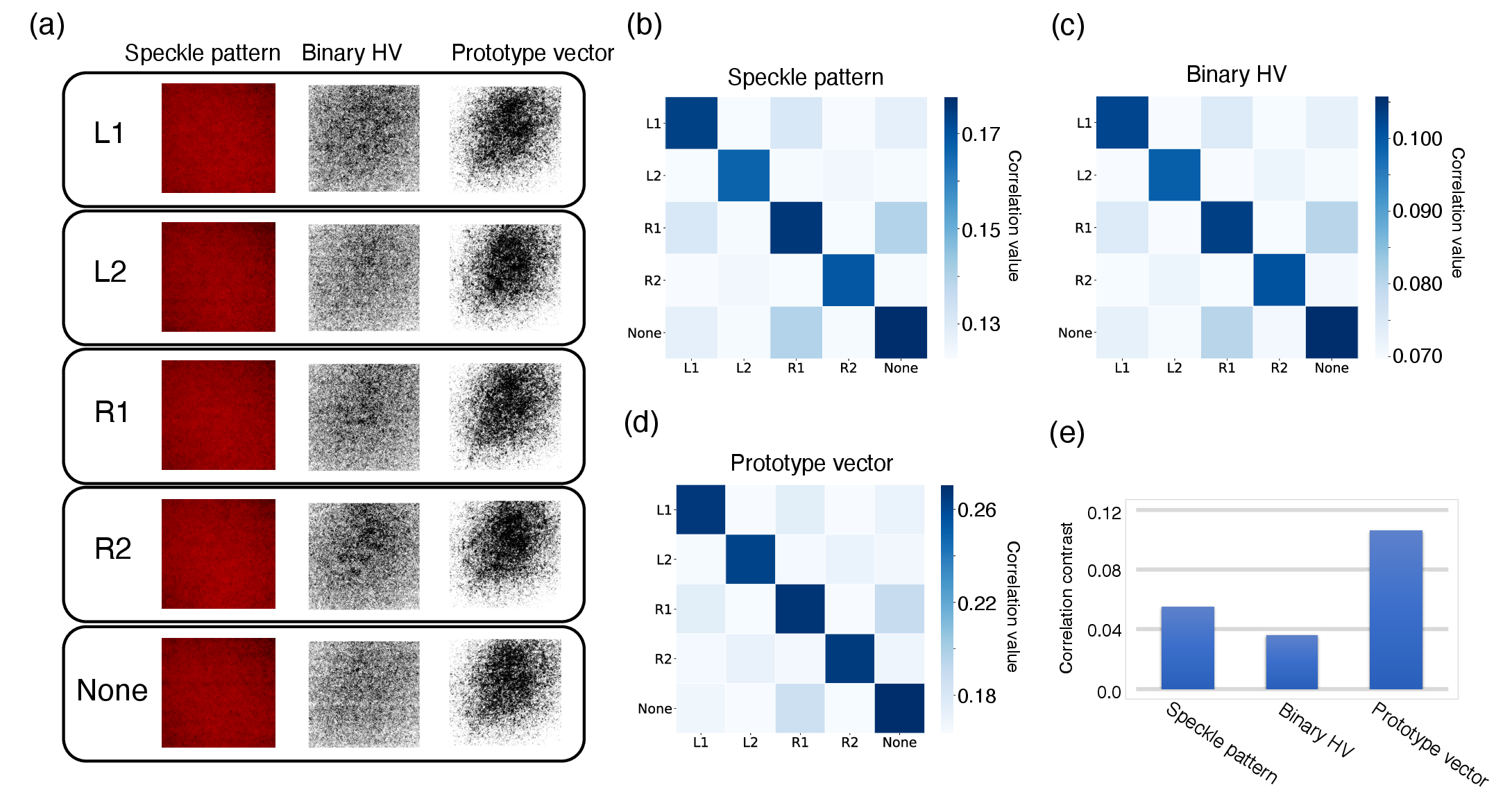}
\caption{\label{fig4} 
(a) Measured speckle patterns, binary HV images, and prototype vector images for each label. 
(b) Correlation matrix $C^{s}$ between speckle patterns. 
(c) Correlation matrix $C^{b}$ between HVs. 
(d) Correlation matrix $C^{p}$ between prototype vectors and binary HVs.  
(e) Correlation contrasts for speckle patterns $\Delta C^{s}$, binary HVs $\Delta C^{b}$, and prototype vectors $\Delta C^{p}$. 
}
\end{figure}

Each speckle pattern was trimmed into an image of approximately 500$\times$500 pixels, flattened, and thresholded to generate binary HVs with approximately 250,000 dimensions. 
Subsequently, the prototype vectors for each class were generated according to Eq.~(\ref{eq_1}). 
Figure~\ref{fig4}(a) shows examples of the speckle patterns, the corresponding HVs, and the corresponding prototype vectors, for each contact position (L1, L2, R1, R2, and None), where the HVs and prototype vectors are reshaped to the same shape as the corresponding speckle images for comparison.
Different speckle patterns were formed for each contact.
To demonstrate this feature, we measured the mean correlation between the speckle patterns,
\begin{eqnarray}
C^{s}_{ll'} = \dfrac{1}{M}\sum_{k=1}^{N_s}\sum_{k'\ne k}^{N_s}\left(
\dfrac{\langle (X_{lk}(i,j)-m^{x}_{lk})(X_{l'k'}(i,j)-m^{x}_{l'k'})\rangle_{ij}
}
{
\sigma^{x}_{lk}\sigma^{x}_{l'k'}}
\right), \label{eq_corr}
\end{eqnarray}
where $X_{lk}(i,j)$ represents the $k$-th speckle pattern labeled by $l \in \{\mbox{L1, L2, R1, R2, None}\}$, $M = N_s(N_s-1)$, and $N_s = 100$.
$(i,j)$ denotes the two-dimensional pixel coordinate in the speckle image. 
$\langle \cdot \rangle_{ij}$ represents the mean with respect to $i$ and $j$.
$m^{x}_{lk}$ and $\sigma^{x}_{lk}$ denote the mean and standard deviation of speckle pattern $X_{lk}(i,j)$, respectively.
The correlation matrix $C^{s}_{ll'}$ is shown in Fig.~\ref{fig4}(b).
The diagonal elements of the matrix 
$C^{s}_{ll}$ (i.e., correlation values between speckle images belonging to 
the same class $(l = l')$) are larger than the non-diagonal elements $C^{s}_{ll'}$ $(l \ne l')$. 
However, $C^{s}_{ll}$ was at most 0.178. 
The maximum contrast between correlation values 
$\Delta C^{s} = \max_{ll'}|C^{s}_{ll}-C^{s}_{ll'}|$ was approximately 0.055.
It is difficult to find any common features among the speckle images belonging to each class. 

We also measured the mean correlation matrix between binary HVs, 
\begin{eqnarray}
C^{b}_{ll'} = \dfrac{1}{M}\sum_{k=1}^{N_s}\sum_{k'\ne k}^{N_s}\left(
\dfrac{\langle (V_{lk}(i)-m^{v}_{lk})(V_{l'k'}(i)-m^{v}_{l'k'})\rangle_i
}
{
\sigma^{v}_{lk}\sigma^{v}_{l'k'}}
\right), \label{eq_corr2}
\end{eqnarray}
where  $V_{lk}(i)$ represents the $i$-th component of the $k$-th sample HV labeled as $l$.
$m^{v}_{lk}$ and $\sigma^{v}_{lk}$ denote the mean and standard deviation of $V_{lk}(i)$, respectively.
The results are shown in Fig.~\ref{fig4}(c).
One can see a similar trend with the result shown in Fig.~\ref{fig4}(b).
The correlation values were low even for $l = l'$, and the correlation contrast $\Delta C^{b} = \max_{ll'}|C^{b}_{ll}-C^{b}_{ll'}|$ for the binary HVs was approximately 0.036.
However, the prototype vectors might contain the information of all HVs belonging to the same class by the bundling operation [Eq.~(\ref{eq_1})] and constitute features representing each class.
To gain further insight into the role of prototype vectors, we computed them with 80 HVs for each class and measured the correlation matrix between the prototype vectors and binary HVs for each class, 
\begin{eqnarray}
C^{p}_{ll'} = \dfrac{1}{N_s}\sum_{k=1}^{N_s}\left(
\dfrac{\langle (PV_{l}(i)-m^{p}_{l})(V_{l'k}(i)-m^{v}_{l'k})\rangle_i
}
{
\sigma^{p}_{l}\sigma^{v}_{lk}}
\right), \label{eq_corr3}
\end{eqnarray}
where  $PV_{l}(i)$ represents the $i$-th component of the prototype vector labeled as $l$.
$m^{p}_{l}$ and $\sigma^{p}_{l}$ denote the mean and standard deviation of $PV_{l}(i)$, respectively.
As seen in Fig.~\ref{fig4}(d), the correlation matrix shows relatively high correlation values for the same class, $l = l'$. 
The correlation contrast $\Delta C^{p} = \max_{ll'}|C^{p}_{ll}-C^{p}_{ll'}|$ was approximately 0.1, which was higher than that for the speckle patterns and binary HVs [Fig.~\ref{fig4}(e)].
This improvement suggests the capability of the prototype vectors to achieve better classifications.

\subsubsection{Classification}
\begin{figure}[ht!]
\centering\includegraphics[width=13cm]{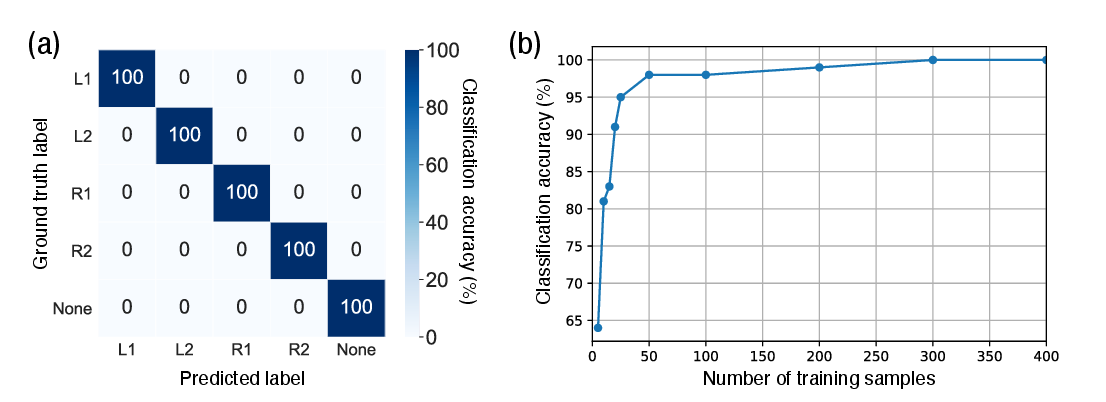}
\caption{\label{fig5} 
(a) Confusion matrix. The number of training samples, $N = 400$. 
(b) Accuracy dependence on the number of training samples $N$.
}
\end{figure}

Here, we discuss the classification of the contact positions. 
During the training phase, five prototype vectors were generated with $N = 80\times5 = 400$ samples.
For the performance evaluation, we used 100 different test samples and generated query HVs. 
The similarity between the query HVs and the prototype vectors was evaluated [Eq.~(\ref{eq_2})], and the contact positions (class labels) were identified. 
The classification results are shown in Fig.~\ref{fig5}(a), exhibiting an accuracy of 100$\%$. 
Figure~\ref{fig5}(b) shows the accuracy dependence on the number of training samples $N$. 
For this simple task, the accuracy exceeded 90$\%$ only when $N > 20$.
This demonstrates the capability of the HDC-based approach to achieve a high classification accuracy using only a few training samples.

%\subsubsection{Comparison with deep learning and Softmax regression}
Another advantage of the proposed HDC-based sensing approach is its straightforward learning method, which does not require an iterative training process using a large number of parameters, as opposed to traditional machine learning. 
We measured the training time of the HDC-based approach as the time required to generate the prototype vectors. 
The time was only 1.24 s for $N = 400$.
For comparison, we also measured the training time, training accuracy, and test accuracy for several other machine-learning models.
Table~\ref{table1} presents the comparison results in terms of the training time and classification accuracy. 
We used Softmax regression, a convolutional neural network (CNN) with a single convolutional layer and max pooling, and a CNN with three convolutional layers and max pooling.
For these computations, a personal computer (Apple Mac mini 2020, OS: macOS 13.2.1, CPU: Apple M1, memory: 16GB) was used.
The training for these machine learning models was unstable because the number of training samples $N$ was limited to 400.
The learning rate was set as 0.0001 to ensure stable training.
The training time was measured as the time in which the accuracy exceeded 90$\%$. 
The training accuracy and test accuracy were measured as the maximum values. 
As shown in Table~\ref{table1}, the training time of the proposed HDC-based approach was significantly shorter than that of the Softmax regression and CNN models, and the test accuracy was slightly higher than those of these machine-learning models.  
The proposed approach reduces the computation burden for the training and enables high classification accuracy with only a limited number of the data samples.

\begin{table}
\begin{center}
 \caption{\label{table1} Comparisons of training time and classification accuracy. 
The number of training samples was set as $N = 400$. 
Softmax regression, a CNN with a single convolutional layer and max pooling [CNN (1 layer)], and a CNN with three convolutional layers and max pooling [CNN (3 layer)] were used in the comparison. 
}
\begin{tabular}{|c|c|c|c|}
\hline
Model & Training time (sec) & Training accuracy ($\%$) & Test accuracy ($\%$)\\
\hline
HDC (this work)  & 1.24 & - & 100\\
\hline
Softmax regression  & 170 & 99.2 & 90\\
\hline
CNN (1 layer)  & 101.3 & 95.8 & 93\\
\hline
CNN (3 layers) & 441.3& 97.3 & 93 \\
\hline
\end{tabular}
\end{center}
\end{table}

\subsubsection{Human-machine interface}
\begin{figure}[ht!]
\centering\includegraphics[width=13cm]{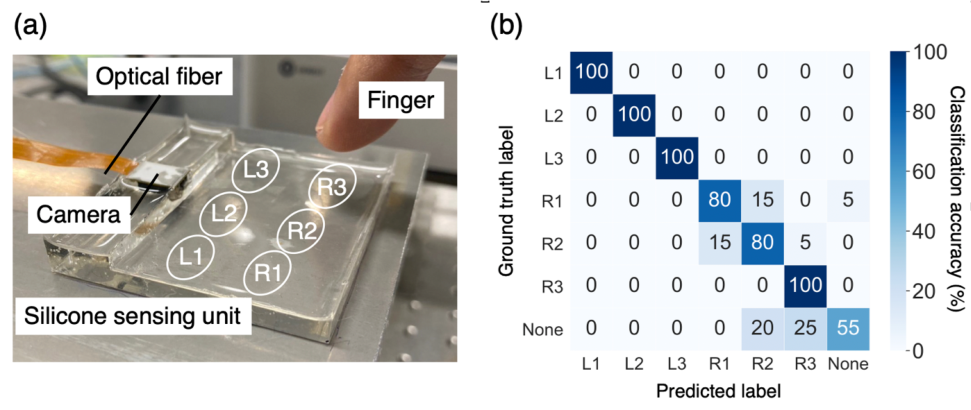}
\caption{\label{fig6} 
(a) Optical touch sensor as an interface device. 
L1, L2, L3, R1, R2, R3, and None can be identified with the HDC-based approach. 
%
%The data samples were collected via the finger, unlike the experiment shown in Fig.~\ref{fig3}. 
%
(b) Confusion matrix.  
}
\end{figure}

In the above demonstration, a solid indenter was used to accurately push the silicone material with the same pressure and direction. 
However, when we utilize the proposed scheme in a human-machine interface, position identification for indentation under various conditions is required.
We investigated whether the sensor could identify contact with a person's finger.
Training samples were collected by repeatedly touching the surface of the sensing unit with a person's index or middle finger [Fig.~\ref{fig6}(a)]. 
The classification results are shown in Fig.~\ref{fig6}(b). 
%
%$N = 400$ samples were used for training. 
%
The accuracy was approximately 87.9$\%$ for 20 test samples for each label. 
The error mainly occurred at positions R1, R2, and R3. 
R1 was confused with R2, whereas None was confused with R2 or R3.
These contact positions are far from the camera, making it difficult to detect the optical signal including the information of the deformation around the contact positions using the camera. 
A straightforward approach to address this issue is to make the silicone material more diffusive by introducing scatterers inside the material, which generate stronger scattering, such that the optical signal containing contact information can be well detected by the camera. 

\subsubsection{Spatial resolution}
\begin{figure}[ht!]
\centering\includegraphics[width=13cm]{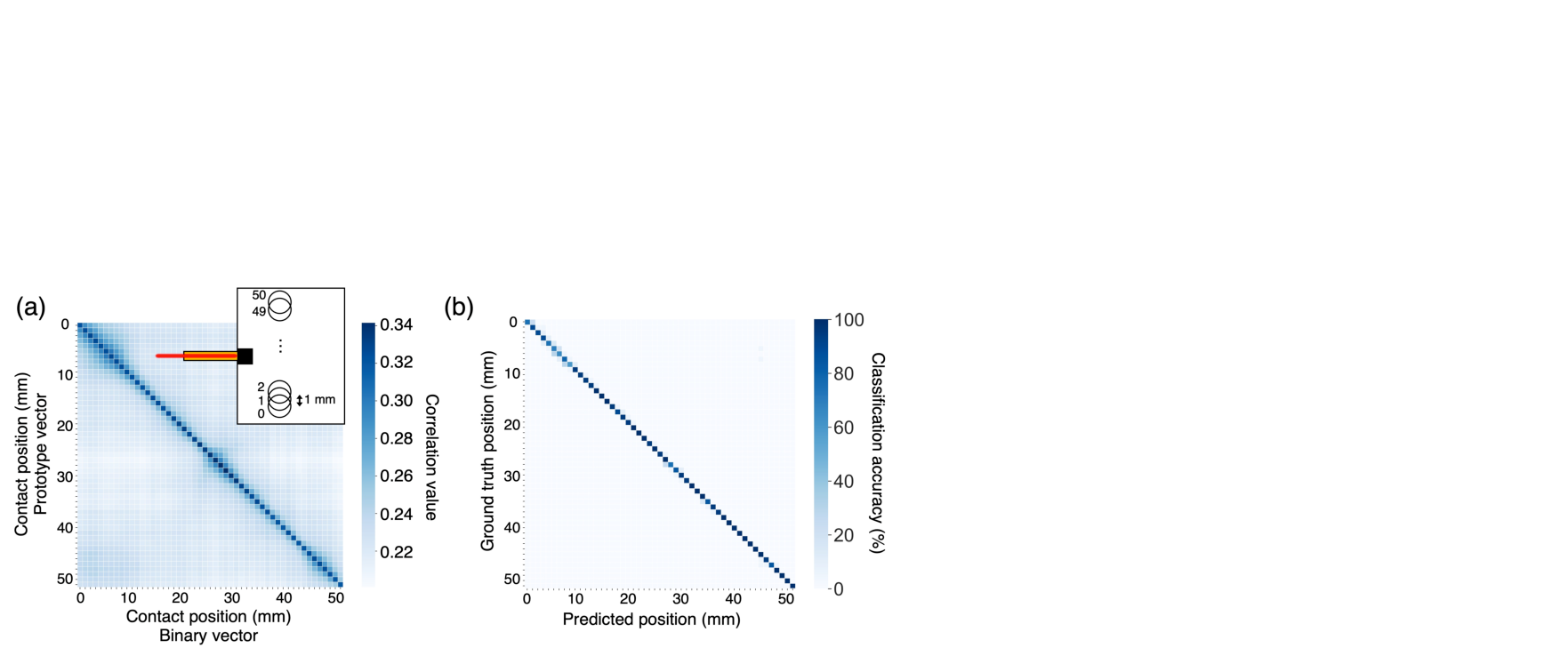}
\caption{\label{fig7} 
Correlation matrix between the prototype vectors and binary HVs. 
The inset in (a) shows the contact positions on the sensing unit. 
(b) Confusion matrix. 
}
\end{figure}

The HDC-based sensing approach does not require the integration of multiple sensors or extensive wiring.
Spatially continuous position sensing is possible via optical scattering. 
To roughly estimate the spatial resolution for identifying the contact positions, we measured the speckle patterns and corresponding HVs formed at certain contact positions.
In this experiment, the contact positions were shifted at a 1-mm interval, and the speckle patterns were measured at 51 contact positions, as labeled as $\{\text{''0''},\text{''1''},\cdots,\text{''50''}\}$ [See the inset in Fig.~\ref{fig7}(a)]. 
The prototype vectors were computed using 80 samples for each position. 
Figure~\ref{fig7}(a) shows the correlation matrix between the prototype vectors and the binary HVs%
The correlation contrast shows that the sensor can identify different contact positions with a 1-mm resolution, which is close to the positioning precision of the indenter used in this experiment.  
Figure~\ref{fig7}(b) shows the identification performance of 51 contact positions at a 1-mm resolution.
The total accuracy was approximately 93$\%$. 
These results suggest the scalability and capability of high-resolution identifications for the proposed speckle-based soft interface. 

\subsection{Adaptive update for varying conditions}
Speckle-based sensing generally causes a stability issue because the speckle patterns are highly sensitive to external environmental variations, including temperature changes and laser fluctuations.  
For example, in our soft interface device, the accuracy was reduced to 32$\%$ 16 days after training due to an environmental change.
To address this stability issue, an adaptive recalibration strategy can be incorporated \cite{Moin:2021aa}. 
This enables updates of the prototype vectors and recovers accuracy without requiring a large number of training samples even under environmental changes. 
%
%For testing, we used the sequential-learning approach proposed in \cite{Moin:2021aa} to update the prototype vectors.
%

\begin{figure}[ht!]
\centering\includegraphics[width=10cm]{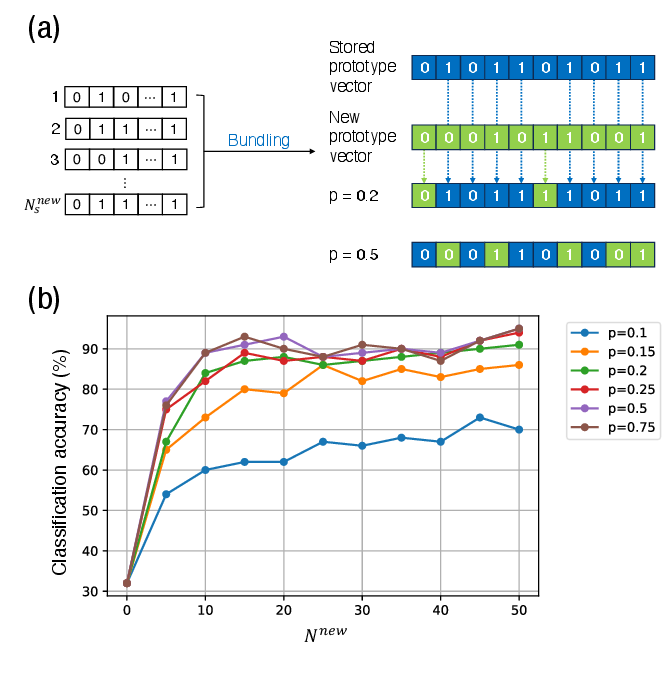}
\caption{\label{fig8} 
(a) Adaptive recalibration scheme based on the update of the prototype vectors.
(b) Classification accuracy vs $N^{new}$.
For $N^{new}/N^{old} =1/8$, the accuracy recovered to 95$\%$ from 32$\%$.
}
\end{figure}

In the update scheme [Fig.~\ref{fig8}(a)], the HVs are acquired in a newly experimental environment, and the new prototype vectors are computed. 
The number of the acquired samples used for the newly computed prototype vectors, $N^{new}$, can be less than the number of samples used for the stored prototype vector, $N^{old}$.
Then, the stored prototype vectors are updated by merging the stored and newly computed prototype vectors with the weight parameter $p$ [Fig.~\ref{fig8}(a)].  
Specifically, an updated prototype vector was made by randomly taking the elements from a newly computed prototype vector with the probability $p$ and replacing the elements of the stored prototype vector with them.
For the proof-of-concept, we measured HVs 16 days after training and updated the prototype vectors by the aforementioned scheme. 
The total number of the acquired samples was set as $N^{new} = N^{new}_sL$, where 
$N_s^{new}$ and $L = 5$ represent the number of the acquired samples for one prototype vector and the number of classes, respectively. 
Figure~\ref{fig8}(b) shows the $N^{new}$-dependence of the classification accuracy.
We measured the accuracy for various values of $p$. 
For the stored (old) prototype vectors, the accuracy was 32 $\%$ but recovered up to 95 $\%$ when $p = 0.5$ or $0.75$ and $N^{new}$ increases to 50, which is 1/8 of $N^{old}$. 
The adaptivity (i.e., how much information on the old environment is forgotten and updated to that on the new environment) can be controlled by the parameter $p$.

\subsection{Robotic finger for tactile sensing}
\begin{figure}[ht!]
\centering\includegraphics[width=13cm]{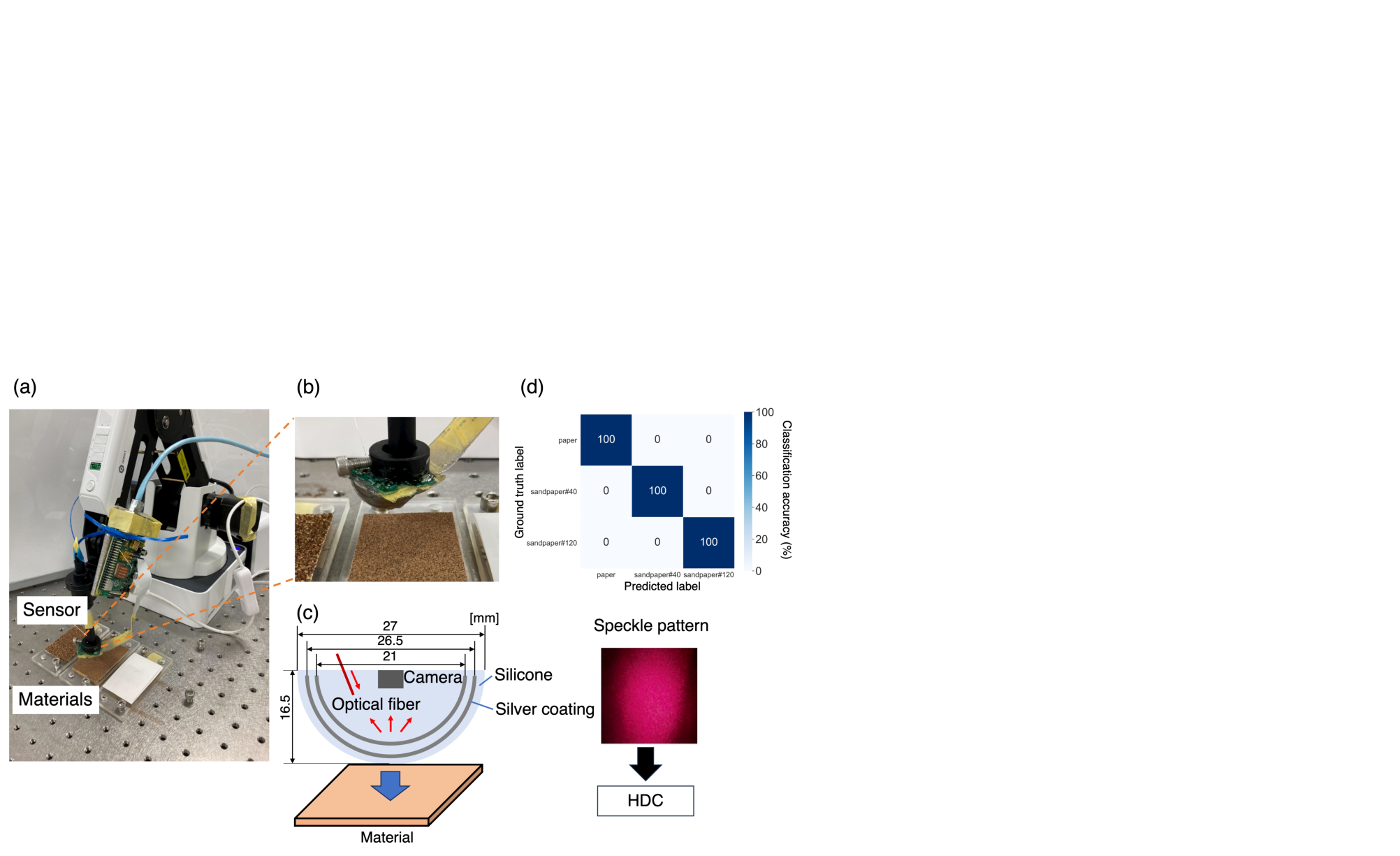}
\caption{\label{fig9} 
(a) Optical tactile sensor mounted as a robotic finger.
(b) Enlarged view of (a). 
(c) Dimensions and structure of the tactile sensor. 
(d) Confusion matrix. 
}
\end{figure}

Finally, we demonstrated that the proposed optical approach can be applied to a robotic finger for tactile identification.
Figures \ref{fig9}(a) and \ref{fig9}(b) show the developed tactile sensor, which is deployed as a robotic finger.  
The simple sensor constitutes a soft silicone elastomer, coupled to an optical fiber, and a compact camera [Fig.~\ref{fig9}(c)].
(See Supplement 1, Section 1D, for further details.)
The tactile sensor differs from a vision-based tactile sensor \cite{7803400} in the sense that markers are not used in the proposed approach. 
Instead, the speckle patterns were measured using the camera because the speckle-based approach allows for high sensitivity \cite{Shimadera:2022aa}.

In the training phase, this sensor was repeatedly touched with three objects, namely, paper, and two sandpapers with grit sizes of 40 and 120, and the training samples were collected.
The grit size of the sandpaper refers to the particle size of the abrading materials embedded in the sandpaper.  
During training, we used 80 samples for each label ($N = 80\times 3 = 240$) and generated the prototype vectors.
The performance was evaluated using 120 test samples.
The results are shown in Fig.~\ref{fig9}(d).
The identification accuracy for touching objects was 100$\%$.

\section{Conclusion}
In this study, we presented an optical approach for HV generation in HDC. 
The proposed approach was applied to a soft interface for contact position identification and to an optical tactile sensor for touching object identification. 
The proposed approach enable fast and efficient cognitive processing without requiring an iterative training process using a large amount of data.  
The sensors can be adaptively recalibrated and maintained high classification accuracy by updating the prototype vectors.
Such computational efficiency and adaptivity contrast with those of traditional techniques using deep learning.

Practically, it is important to increase the number of contact positions in the identifications for the interface device and touching objects for the tactile sensor.
This can be overcome by increasing the number of optical fibers and/or introducing scatterers inside the elastomer for increased diffusivity.  
In addition, the HDC-based approach can be extended to infer continuous values using a recently developed algorithm \cite{9586284}.

The proposed approach enables the resolution of various issues inherent in speckle-based sensing techniques and can be utilized for low-latency and adaptive in-sensor optical processing.

%\section{Appendix}

%\begin{equation}
%J(\rho) =
% \frac{\gamma^2}{2} \; \sum_{k({\rm even}) = -\infty}^{\infty}
%	\frac{(1 + k \tau)}{ \left[ (1 + k \tau)^2 + (\gamma  \rho)^2  \right]^{%3/2} }.
%\end{equation}

\section*{Acknowledgments}
This work was partly supported by JSPS KAKENHI (Grant Nos. JP22H05198,JP22K18792,JP22H01426) and JST PRESTO (Grant No.~JPMJPR19M4).

\end{document}